# NoLiTiA: An Open-Source Toolbox for Nonlinear Time Series Analysis

*Immo Weber*[*][a], *Carina R. Oehrn*[a,b]

[a] Department of Neurology, Philipps-University of Marburg, Marburg, Germany

[b] Center for Mind, Brain and Behavior (CMBB), Philipps-University Marburg, Marburg, Germany

* Corresponding author: immo.weber@staff.uni-marburg.de

# Abstract

In many scientific fields like e.g. neuroscience, climatology or physics, complex relationships can be described most parsimoniously by nonlinear mechanics. Despite their relevance, many scientists still apply linear estimates in order to evaluate complex interactions.

This is partially due to the lack of a comprehensive compilation of nonlinear methods. Available packages mostly specialize in only one aspect of nonlinear time-series analysis and most often require some coding proficiency to use.

Here, we introduce NoLiTiA, a free open-source MATLAB toolbox for nonlinear time series analysis.

In comparison to other currently available nonlinear packages, NoLiTiA offers 1) an implementation of a broad range of classic and recently developed methods, 2) an implementation of newly proposed spatially and time-resolved recurrence analysis and 3) an intuitive environment accessible even to users with little coding experience due to a graphical user interface and batch-editor.

The core methodology derives from three distinct fields of complex systems theory, including dynamical systems theory, recurrence quantification analysis and information theory. Besides established methodology including estimation of dynamic invariants like Lyapunov exponents and entropy-based measures, such as active information storage, we include recent developments of quantifying time-resolved aperiodic oscillations. In general, the toolbox will make nonlinear methods accessible to the broad scientific community engaged in time series processing.

## 1. Introduction

In science, researchers usually strive to explain natural phenomena by the most parsimonious way possible. Based on experimental evidence, the scientist formulates a model of how the data was generated and successively tests it. A model is deemed fitting if it is able to make prediction with an acceptable accuracy. For simple relationships like e.g. of force and acceleration in classic Newton mechanics, linear models offer the best fit. In these models, small variations of the inputs lead to small differences in the output observables. All of the remaining variance left unexplained by the model is thought to be generated by noise, either due to the circumstances of the measurement, or due to some inherent stochasticity of the underlying system (e.g. Brownian motion). While linear models capture some phenomena in nature, they insufficiently explain more complex systems e.g. as studied in neuroscience (Faure and Korn 2001), climatology (Ghil et al. 1991), geophysics (Lovejoy et al. 2009) or genetics (Mazur et al. 2009). Complex systems are often characterized by nonlinear relationships, with unique features not observed in linear systems. Here, small input variations may lead to drastic differences in a system's future behaviour e.g. observed in neurons during generation of action potentials (Fitzhugh 1960). Also, nonlinear systems often produce emergent behavior, e.g. swarm behavior (Garnier et al. 2007), where its collective dynamics cannot trivially be described by a sum of its parts (principle of superposition). Whereas linear models typically need stochastic components to capture complex relationships, nonlinear models - e.g. for predator-prey relationships (Danca et al. 1997) - often suffice with simple deterministic equations, thereby increasing their predictive value. The field of complex systems theory is a synthesis of many research areas, including the study of dynamical systems, information theory and recurrence analysis (Nicolis and Rouvas-Nicolis 2007). With increasing computational capabilities of modern computers, these methods become more easily applicable and may thus become complementary to standard linear techniques. With the advent of machine learning approaches, nonlinear statistics might prove to be valuable features complementing linear statistics, as has been shown e.g. in studies to distinguish neurological from healthy subjects (Hosseinifard et al. 2013; Yuan et al. 2011; Pezard et al. 2001). Further, methods from signal-processing may produce spurious results for nonlinear signals, as has been recently demonstrated for phase-amplitude coupling (Lozano-Soldevilla et al. 2016; Gerber et al. 2016). Up to this point a comprehensive toolbox combining methods from the wide field of complex systems theory within a unified framework is lacking. Thus, we developed NoLiTiA, a free, open-source MATLAB toolbox for nonlinear time series analysis. The toolbox covers established and novel methods from three distinct fields of complex systems theory: dynamical systems theory (Kantz and Schreiber 2004), recurrence quantification analysis (Eckmann, J-P., S. Oliffson Kamphorst, and David Ruelle. 1987) and information theory (Shannon and Weaver 1949). Our objective is to provide an easily accessible, intuitive toolbox to a broad scientific community.

In the following sections we will first give a short introduction to the topics covered by the toolbox, explain the general workflow and implementation, and finally validate the core methodology and present some example applications.

## 2. Core Methodology

The full documentation of all functions and parameters can be found in the toolbox manual (see supplement).

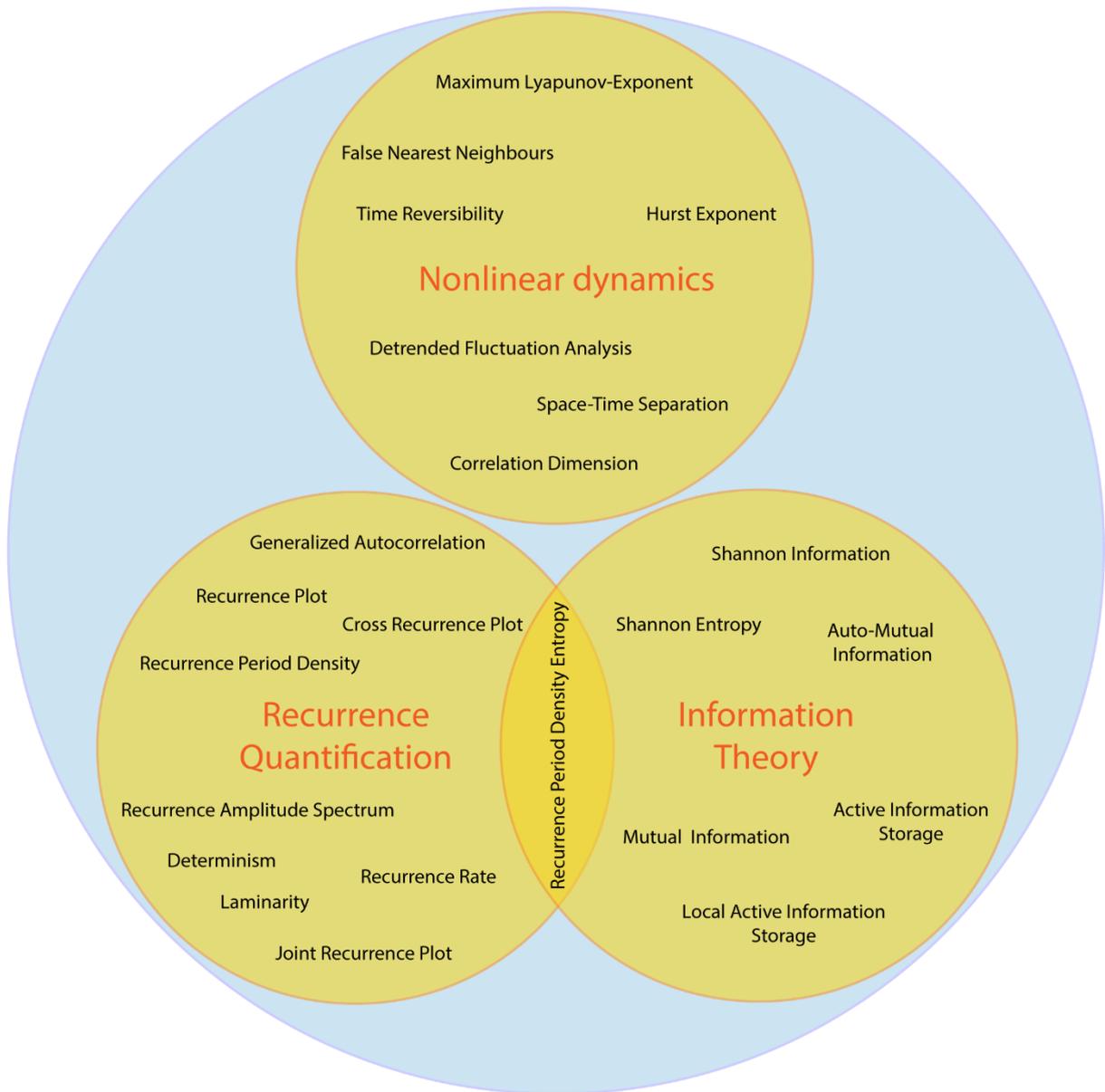

**Figure 1:** Topics and measures covered by NoLiTiA.

*Dynamical Systems Theory*

The field of dynamical systems theory includes methods to characterize the temporal evolution of phase-space observables (states), i.e. the combination of all independent variables uniquely determining a state in multidimensional systems (Kantz and Schreiber 2004). An intuitive example is the pendulum, of which the current state is uniquely determined by its momentary position and acceleration. Thus, its phase-space is two dimensional and in the case of no friction forms an unstable periodic orbit, i.e. closed trajectory in response to small perturbations of its resting position. With friction, phase-space trajectories spiral into a fixed point at the resting position of the pendulum. In the case of a driven pendulum, the phase-space also forms a closed trajectory, which attracts nearby trajectories induced by small perturbations. In contrast to the undriven pendulum with no friction it is thus a stable periodic orbit, as it is resistant to small perturbations. Both, the fixed point of an undriven pendulum with friction and the closed trajectory of a driven pendulum are examples of so-called attractors and determine the systems qualitative behavior. As the name implies, attractors attract system states in its near vicinity and keep its dynamics bounded to a specific region in phase-space. Nonlinear systems may possess very complex attractors which are composed of an indefinite number of fixed points and periodic orbits. Using methods from dynamical systems theory, it is possible to characterize such attractors and estimate short and long term behavior of nonlinear systems, e.g., how fast initially similar states diverge over time in phase-space. This allows crucial insights into the temporal stability of the underlying system and furthermore makes it possible to distinguish random from chaotic or highly complex but deterministic behavior. By studying the dynamic skeleton of the system, i.e. its fixed points and (unstable) periodic orbits, it is also possible to control its behavior to a certain extent (Schöll 2010; Bielawski et al. 1994; Zhu et al. 2019). For most methods from dynamical systems theory, it is necessary to have a phase-space representation of the measured time series. Although, in most experimental setups it is not possible to measure all state variables simultaneously, under certain conditions, Takens' delay embedding theorem (see section below) guarantees the reconstruction of a topologically equivalent phase-space from univariate time series (Takens 1981). The implemented methods include e.g. testing for nonlinearity (Schreiber and Schmitz 1997), phase-space reconstruction (Takens 1981), topological complexity measures (correlaton dimension, Grassberger and Procaccia 1983, false nearest neighbours, Kennel et al. 1992), nonlinear prediction (Ragwitz and Kantz 2002), measures for detecting temporal instability (Lyapunov exponents, Kantz 1994; Rosenstein et al. 1993) and for detecting unstable periodic orbits (So et al. 1996).

*Test for Nonlinearity*

Many methods from classic signal processing like Fourier analysis, but also newer methods like phase-amplitude coupling rely on the assumption of linearity. NoLiTiA utilizes a test for nonlinearity based on the generation of surrogates. According to Schreiber and Schmitz (1997) quantification of temporal asymmetry of a time series X with data points $x_t$ is most sensitive to distinguish linear from nonlinear processes:

$$Q_t(\Delta t) = \langle (x_t - x_{t-\Delta t})^3 \rangle \qquad\qquad ( 1 )$$

with <..> indicating the average. The measure quantifies the sharpness of a signal's transition backwards in time. Smoother signals are more likely to be generated by linear processes. However, to exclude a stochastic origin, results need to be compared to surrogates. The time-inversion statistic $Q_t$ is first calculated for the original data and then for a specified number of surrogates. A two-sided Z-statistic is then calculated and compared to the distribution of surrogates. If the absolute Z-score is smaller than 1.96 the test is significant at an alpha-level of 5%. The toolbox provides several algorithms for surrogate data generation to test the null hypothesis of test data being generated by a linear process. The implemented algorithms include shuffling of time points, shuffling of data segments, phase-randomization and amplitude-adjusted phase-randomization. Briefly, the latter algorithm for generating surrogates is implemented as follows: 1) original data is Fourier transformed, 2) phases of Fourier transformed data are randomized, 3) the inverse of the Fourier transform is performed on phase randomized data. After these steps the following four steps are repeated iteratively: 4) rank sort surrogate data according to original data 5) calculate Fourier transform of surrogates, 6) exchange surrogate amplitudes with amplitudes of original data, 7) calculate the inverse of Fourier transform (Schreiber and Schmitz 1996). The test may be invoked by calling the nta_timerev function. At least the surrogate method (default: phase-randomization) and the number of surrogates (default:100) should be specified.

*Phase-Space Reconstruction*

Embedding of univariate time series in phase-space is implemented using Takens' delay embedding theorem (Takens 1981). By time-shifting a univariate time series x(t) d times by a factor τ, the time series can be embedded in a d-dimensional phase-space:

$$\boldsymbol{x}_t^{d_x} = [x_{t-(d_x-1)\tau}, x_{t-(d_x-2)\tau}, \dots, x_{t-\tau}, x_t]^{TP} \qquad (2)$$

where TP indicates the transpose and $x_t$ being the d-dimensional state vector at time t.

The phase-space may be reconstructed using the nta_phasespace function:

```
load('lorenz10000.mat')
data            =       x;
cfg             =       [];
cfg.dim         =       3; % embedding dimension
cfg.tau         =       9; % embedding delay
[ results ]     =       nta_phasespace( data,cfg );
```

The two embedding parameters cfg.dim (=d) and cfg.tau need to be specified to get reasonable results. The phase-space matrix (time x embedding dimension) is returned in the field embTS of the results structure array. As phase-space embedding is a key procedure in nonlinear time series analysis, nta_phasespace is called as a subfunction by most other routines in the toolbox.

Despite the possibility to define the necessary parameters for phase-space reconstruction *ad-hoc*, both parameters dim and τ may be automatically optimized based on whether data is generated by a stochastic or deterministic process. The deterministic approach involves the method of false nearest neighbours for dimensionality optimization and auto-mutual information for τ optimization (Kennel et al. 1992; Cao 1997, see section 2.3.). The stochastic approach co-optimizes dim and τ using the Ragwitz prediction criterion for Markov-models (Ragwitz and Kantz 2002) (see section Nonlinear Prediction).

Embedding optimization may be invoked by calling the nta_optimize_embedding function.

*False Nearest Neighbours*

For the reconstruction of the phase-space, the dimension d may be chosen *ad-hoc* or optimized with regard to some advantageous criteria. One possibility is to optimize d with regard to the topology of the phase-space using the false nearest neighbours approach introduced by Kennel et al. (1992). The algorithm (nta_fnn) quantifies the number of false neighbouring points of a time series X in a reconstructed phase-space which arise due to an insufficient embedding. If the embedding dimension is chosen too low, points which should be far away on the attractor may become direct neighbours, due to them being projected into each other's vicinity. A toy example would be two points in a Cartesian two-dimensional coordinate system, which have the same x-coordinate but different y-coordinates. Projecting both points on the x-axis, i.e. reducing the dimension, would results in both points being on top of each other, or in other words, become false neighbours. On the other hand, projecting both points into the next higher dimension does not increase their distance to each other. The false nearest neighbours approach embeds the time series successively into higher dimensions and counts the reduction of neighbours, which arise due to projections:

$$\text{FNN}(d) = \sum_{\wedge} \frac{\theta\left(\frac{\left|x_{i+1} - n(x_i)_{j+1}\right|}{\left\|\mathbf{x}_i^d - \mathbf{n}_{xi}^d\right\|} - \text{Rtol}\right)}{\theta\left(\sqrt{\left(\left\|\mathbf{x}_i^d - \mathbf{n}_{xi}^d\right\|\right)^2 - \left(\left|x_{i+1} - n(x_i)_{j+1}\right|\right)^2} - \text{SD}(X) * \text{Atol}\right)} \qquad (\,3\,)$$

with θ being the Heaviside-step function, i being the temporal index of x, j being the temporal index of n, $n(x_i)$ indicating the next neighbour of $x_i$, ||..|| indicating the maximum norm, Rtol being a distance threshold, Atol being a loneliness threshold and SD indicating the standard deviation of the time series. The thresholds correct for points in phase-space, which are either to sparse to have direct neighbours or which are too close to the borders of the attractors. The lowest dimension, for which the percentage of nearest neighbours drops to zero is a reasonable choice for embedding (Kennel et al. 1992). Besides τ a range of embedding dimensions should be specified to test for (default: 1-9).

*Nonlinear Prediction*

While the false nearest neighbours algorithm is well suited for determining the embedding dimension of deterministic systems it may return spurious results if the underlying process is generated by a stochastic process. Ragwitz et al. proposed the application of a simple nonlinear predictor (Farmer and Sidorowich 1987) to embed stochastic processes with the Markov property, i.e. processes with finite memory (Ragwitz and Kantz 2002). Here d-dimensional states $x_t$ at each time point t are predicted by averaging the iterates of all closest spatial neighbours of a neighbourhood U within a distance $\varepsilon$ (function: nta_ragwitz):

$$\hat{\mathbf{x}}_{t+1}^{d_x}(d, \tau) = \frac{1}{\left|U_\varepsilon(\mathbf{x}_t^{d_x})\right|} \sum_{\mathbf{x}_{t-\Delta t}^{d_x} \in U_\varepsilon(\mathbf{x}_t^{d_x})} \mathbf{x}_{t-\Delta t+1}^{d_x} \qquad (\ 4\ )$$

The root mean squared prediction error (RMSPE) is subsequently calculated.

$$\text{RMSPE}(d, \text{tau}) = \sqrt{\frac{\sum_{t=1}^{N}\left(\hat{\mathbf{x}}_{t+\Delta t}^{d_x} - \mathbf{x}_{t+\Delta t}^{d_x}\right)^2}{N}} \qquad (\ 5\ )$$

This procedure is repeated for all query points N for each combination of embedding parameters dim and $\tau$. The combination of embedding parameters for which the RMSPE is minimum is suggested for further usage. The same approach has been implemented e.g. in the TRENTOOL and JIDT software packages for the estimation of information theoretic measures (Lindner et al. 2011; Lizier 2014). Possible configuration parameters include the range of embedding dimensions (default: 2-9) and delays (default: 10-100 % autocorrelation time).

*Correlation Dimension*

The phase-space representation of nonlinear processes, i.e. the attractor often possesses a scale invariance of its geometrical topology. This so called fractality commonly serves as an indicator for the complexity of the analysed system (Wiltshire et al. 2017; Shekatkar et al. 2017; Hoang et al. 2019). It can be quantified by estimating the correlation dimension D2 (Grassberger and Procaccia 1983) defined by:

$$C(\varepsilon) \sim \varepsilon^{D2}, \qquad (6)$$

where C(ε) counts all pairs of states $x_i$ and $x_j$ on the attractor which are closer than a distance ε:

$$C(\varepsilon) = \frac{2}{N(N-1)} \sum_{i=1}^{N} \sum_{j=i+1}^{N} \Theta\left(\varepsilon - \left\| \mathbf{x}_i - \mathbf{x}_j \right\| \right) \qquad (7)$$

with N being the total number of points, θ being the Heaviside-step function and ||..|| indicating some distance norm.

For chaotic systems, a definite fractal dimension can be quoted, if an estimate of D2 can be found for a reasonable scaling range and if this estimate is invariant for a reasonable number of embedding dimensions higher than the order of the process (Kantz and Schreiber 2004). According to Theiler (1986) temporal neighbours in phase-space have to be excluded. For this so-called Theiler window, one may choose twice the autocorrelation time, i.e. the first zero crossing of the autocorrelation function (Theiler 1986). Another option is to use a so called space-time separation plot (Provenzale et al. 1992), which allows to depict spatial distances in phase-space as a function of temporal distances (function: nta_spacetimesep, also see Figure 9b). Estimation of D2 may be invoked by using the nta_corrdim function.

It is calculated by estimating the slope of a line fitted to the logarithm of C as a function of log(ε):

$$\log C(\varepsilon) = D2 \log \varepsilon \qquad\qquad ( 8 )$$

Among the most important parameters to specify is the range of distances ε to test (default: 1-100 % of attractor diameter).

*Maximum Lyapunov Exponent*

A characteristic feature of nonlinear systems is the exponential divergence of similar states in phase-space (Kantz 1994; Rosenstein et al. 1993):

$$\delta_{\Delta t} \propto e^{\lambda t} \delta_{t_0}, \qquad\qquad ( 9 )$$

with $\delta_{t0}$ indicating the initial distance of two states, $\delta_{\Delta t}$ indicating the distance of the two states after t time steps and λ being the largest Lyapunov exponent, i.e. the divergence rate. A large positive Lyapunov exponent indicates a very unstable system, where small differences of states, e.g. introduced by external perturbances may quickly lead to drastic changes of the temporal behavior. A negative exponent is related to dissipative, i.e. converging dynamics. Estimation of Lyapunov exponents have e.g. been used to explain and even predict seizures in epilepsy (Usman et al. 2019; Aarabi and He 2017; Lehnertz and Elger 1998). In the context of signal processing it has been shown, that the Lyapunov exponent is also relevant for the estimation of other measures like Granger-causality, as systems in which λ approaches zero from below have very unstable autoregressive representations (Friston et al. 2014).

The toolbox implements two algorithms for estimating the maximum Lyapunov exponent:

Kantz' algorithm (Kantz 1994):

$$\delta(\Delta t) = \frac{1}{N} \sum_{t_0=1}^{N} \ln \left( \frac{1}{|U(\mathbf{x}_{t_0})|} \sum_{\mathbf{x}_t \in U(\mathbf{x}_{t_0})} ||\mathbf{x}_{t_0+\Delta t} - \mathbf{x}_{t+\Delta t}|| \right) \qquad\qquad ( 10 )$$

Rosenstein's algorithm (Rosenstein et al. 1993):

$$\delta(\Delta t) = \frac{1}{N} \sum_{t_0=1}^{N} \ln\left(\left\|\mathbf{x}_{t_0+\Delta t} - \mathbf{x}_{t+\Delta t}\right\|\right),\qquad(\,11\,)$$

The difference between both algorithms is that Rosenstein's algorithm only utilizes the closest neighbour of each point, while Kantz' algorithm takes the average of all neighbours within a given neighbourhood-size. Both, however, yield very similar results (Dingwell 2006).

Similar to the correlation dimension an estimation of the maximum Lyapunov exponent for chaotic systems should be invariant for succeeding embedding dimensions of at least the order of the underlying process or else it cannot be concluded to be a property of an invariant attractor of a deterministic system (Kantz and Schreiber 2004). The function nta_lya may be used to estimate the largest Lyapunov exponent.

It is calculated similar to the estimate of the correlation dimension by fitting a straight line to the logarithm of $\delta_{\Delta t}$ as a function of t. Important parameters to specify are the number of temporal iterations $\delta_{\Delta t}$ (default: 10) and, in the case of the Kantz' algorithm, the neighbourhood-size U (default: 5 % of attractor diameter).

*Unstable Periodic Orbits*

In nonlinear systems unstable periodic orbits form its skeleton, thus determining its overall dynamics. By detecting these periodic orbits it is possible to control the systems long term behavior by applying a feedback control scheme (Schöll 2010; Bielawski et al. 1994; Zhu et al. 2019). The function nta_upo may be used to determine the location of any period one orbits in a two-dimensional phase-space. The algorithm implemented in NoLiTiA is the one proposed by So et al. (1996). Briefly, the algorithm applies a transformation procedure, where it locally linearizes the phase-space and projects any points in its vicinity on the nearest fixed point, while simultaneously dispersing any unrelated points far away. The procedure may be repeated n times for maximal efficiency. A surrogate approach using the same routines as described for the nonlinearity testing may be applied to test for statistical significance of any possible fixed points. For one-dimensional maps plotted in two-dimensional phase-space the fixed points lie on the intersection of the first diagonal and the attractor. Thus, only the first diagonal needs to be tested statistically. Most importantly, the user should specify the number of transformations to perform (default: 100) and whether and if how many surrogates should be generated for statistical testing (default: 0).

*Recurrence-based Measures*

Recurrence-based measures are derived from the field of dynamical systems theory and exploit the neighbourhood-relationships of states by reducing arbitrarily high dimensional phase-spaces to two dimensional recurrence matrices M (Eckmann et al. 1987):

$$M_{t,t+\Delta t} = \Theta\big(\varepsilon - ||\mathbf{x}_t - \mathbf{x}_{t+\Delta t}||\big), \qquad\qquad ( \ 12 \ )$$

where $\theta$ indicates the Heaviside-step function, $||..||$ indicates a distance norm (Euclidean norm), $\varepsilon$ is the neighbourhood-size in phase-space and $x_t$ is the phase-space vector at time t. As the name suggests each black dot represents one recurrence in time (Figure 2). Depending on the local dynamics of the system, the recurrence plot shows different motifs. Parallel diagonal lines are a hallmark of periodicity and determinism. In contrast, vertical lines appear due to laminar, i.e. unchanging behavior. White corners may appear due to slow drifts i.e. non-stationarity and isolated dots most often indicate stochastic behavior (Eckmann et al. 1987).

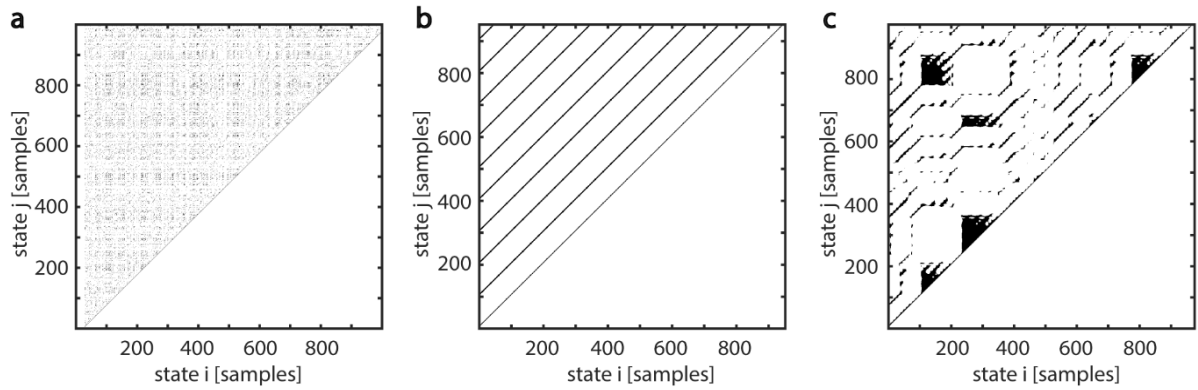

**Figure 2:** Examples for recurrence plots with different motifs. a) White noise, b) periodic signal with a period of 100 samples, c) Lorenz-system (see section 4 for an explanation of the Lorenz-system). X and Y axis represent temporal order of states. Each black dot indicates a recurrence of state i at time j. Figure adapted from Weber and Oehrn (2021).

The recurrence matrix as well as several derived measures like determinism (Marwan and Kurths 2002), laminarity (Marwan et al. 2002), generalized autocorrelation (Romano et al. 2005) or recurrence period density entropy (Little et al. 2007) may be calculated using the nta_recurrenceplot function. An important parameter to specify is either the neighbourhood-size ε (default: 0) or the recurrence rate, which automatically determines the fill-rate of the recurrence plot in percent of the total possible fill-rate, by adjusting individual neighbourhood-sizes for each point in phase-space (default: 5 %).

Aside from classic recurrence quantification analysis measures, as well as bivariate extensions (joint (nta_jointrecurrenceplot) and cross recurrence plots (nta_crossrecurrenceplot, Marwan and Kurths 2002; Romano et al. 2004), two recently proposed measures are implemented: time- and scale-resolved recurrence amplitude analysis (Weber and Oehrn 2021).

*Spatially- and time-resolved recurrence period analysis*

Oscillatory phenomena can be abundantly observed in many scientific fields. Examples are the El-Nino-Southern Oscillation in climatology (Timmermann et al. 2018), the Belousov-Zhabotinsky reaction in chemistry (Hudson and Mankin 1981), predator-prey relationships in biology (Danca et al. 1997) or electrophysiological brain activity in neuroscience (Buzsáki and Draguhn 2004). Oscillatory phenomena are often analysed by means of spectral analysis methods like Fourier- or Wavelet analysis (Muthuswamy and Thakor 1998; Oehrn et al. 2018). A major property of these methods is the assumption, that the time series on which they are applied to can be best characterized by a superposition of sine waves. However, many natural oscillations are non-sinusoidal and would be imperfectly represented by standard spectral methods (Philander and Fedorov 2003; Cole and Voytek 2017). Application of these methods results in spurious harmonics in the frequency domain and might even lead to false positive observations of derived methods, like phase-amplitude coupling (Lozano-Soldevilla et al. 2016; Gerber et al. 2016). Another basic assumption when applying classic methods is a certain periodicity of the analysed signal. Studies frequently observe oscillatory behavior which is not perfectly periodic, but nearly periodic or quasi-periodic (Yousefi et al. 2018; Thompson et al. 2014; Ko et al. 2011; Li et al. 2011). Such signals are represented by broad peaks in the frequency domain, which might even mask adjacent frequency components. Over the last years methods from recurrence analysis have become an alternative to classic spectral methods, especially when analysing quasi-periodic (Zou et al. 2008) or non-sinusoidal (Little et al. 2007) oscillatory activity. In the following, we will briefly describe the estimation procedure of the time-resolved recurrence amplitude spectrum (TRAS) and scale-resolved recurrence period spectrum (SRPS).

A d-dimensional state $x_t$ is defined to be recurrent if, after $\Delta t$ time steps, it is within a neighbourhood $U_\varepsilon$ of $x_t$ (Eckmann et al. 1987):

$$x_{t+\Delta t}^{d_x} \in U_\varepsilon(\mathbf{x}_t^{d_x}) \qquad\qquad (\,13\,)$$

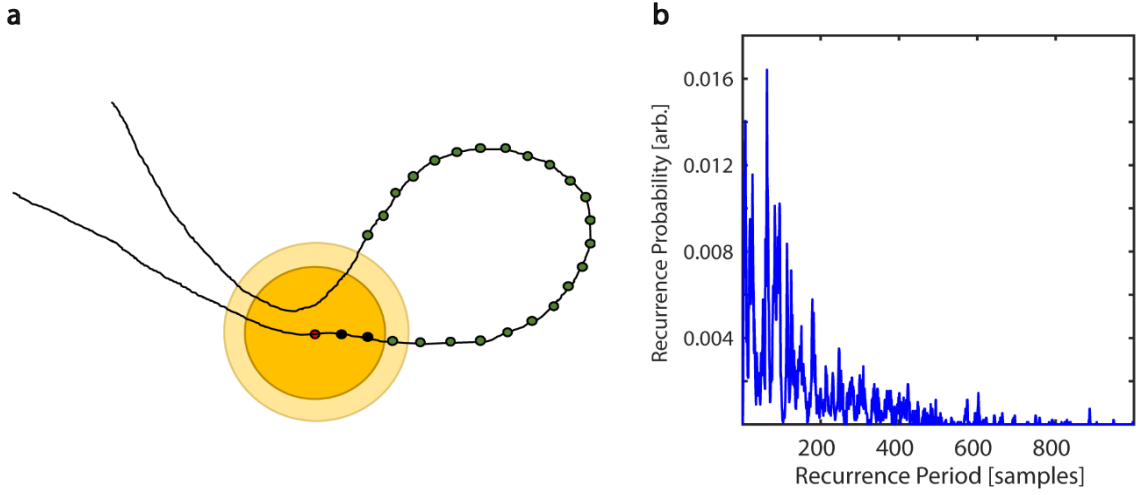

**Figure 3:** Recurrence of states. a) The yellow circles indicate neighbourhoods of different sizes of the reference point (red). The dotted line indicates the temporal flow, while the green and black dots represent succeeding states. In this example, it takes t=25 samples for the reference point to leave and re-enter the smaller neighbourhood (dark yellow), b) Recurrence probability as a function of recurrence periods. Figure adapted from Weber and Oehrn (2021).

For the limit of $\varepsilon \rightarrow 0$ $x_t+\Delta t$ is defined to be periodic with period $\Delta t$. Based on the methods of close returns (Lathrop and Kostelich 1989) Little et al. (2007) calculated the recurrence time T of any closest temporal neighbour $x_t + \Delta t$ of $x_t$ within a spatial neighbourhood $U_\varepsilon$ as the difference:

$$T = (t + \Delta t - \rho) - (t + \gamma),$$ ( 14 )

where $\gamma$ is the difference in samples between $x_t$ and $x_t$ first leaving $U_\varepsilon$. $\rho$ is the sample difference between $x_t$ reentering $U_\varepsilon$ and $x_t+\Delta t$ (Figure 3a). This is equivalent to calculating the number of states between vertical line segments starting at the main diagonal in M (equation 12, see also Figure 2). After repeating this procedure for all state vectors, one can calculate a histogram R(T) with bin size equal to 1 sample and a number of bins equal to the longest recurrence time $T_{max}$. By normalizing R(T) by the total number of recurrences, one gets the recurrence probability density (Little et al. 2007) (Figure 3b). In the following, we will refer to this measure simply as recurrence probability. To account for the high number of short period recurrences in noisy data, P(T) can be calculated for a predefined range of T, i.e. specifying a $T_{min}$ and $T_{max}$.

$$P(T) = \frac{R(T)}{\sum_{i=T_{min}}^{T_{max}} R(i)} , T = T_{min} \dots T_{max} \tag{15}$$

The recurrence probability thus measures the probability of a recurrence occurring after T time steps.

As recurrent states form closed trajectories in phase-space, their energy is contained within the phase-space volume. Thus, a reasonable approximation of the recurrent amplitude of a specific frequency is to estimate the maximum phase-space diameter and weight it by the respective recurrence probability of this frequency:

$$\bar{a}_{diam}(T) = \sum_{k=1}^{q} max \left\| \mathbf{x_i} - \mathbf{x_j} \right\| * q^{-1}, \forall i, j \in n \tag{16}$$

Where q is the number of recurrences per T and n is the number of samples per recurrence (see Figure 3a).

The estimation of P(T) is highly dependent on ε. A choice of ε too small would result in many empty neighbourhoods and thus in a high degree of statistical errors. If ε is chosen too large recurrences are not local anymore and recurrence periods are underestimated. To avoid ambiguity and to effectively eliminate the parameter ε one may calculate P(T) over a wide range of values, resulting in a scale-resolved recurrence period spectrum (SRPS).

$$P(T, \varepsilon) = \frac{R(T, \varepsilon)}{\sum_{i=T_{min}}^{T_{max}} R(i, \varepsilon)} , T = T_{min} \dots T_{max} \tag{17}$$

In NoLiTiA, we calculate SRPS as a function of T and ε. ε may be defined in percent of the standard deviation of the analysed time series. In the SRPS one would expect to find three regions of interest depending on ε (e.g. see Figure 4a). For very small ε the SRPS is governed by statistical errors and a uniform distribution across all T. At the crossing of the noise level, multiples of the true recurrence periods might appear, due to the recurrent points slightly missing the neighbourhood of the reference point. For very high ε the distribution of P(T) is heavily shifted to small T with only few points leaving and re-entering any neighbourhood, with the extreme case of a neighbourhood-size fully engulfing the phase-space. If the time series contains any oscillatory activity, slowly shifting but continuous spectral peaks occur in the intermediate range of ε. The shifting occurs, as the increasing neighbourhood-size succeedingly engulfs more and more adjacent samples, which leads to an underestimation of the recurrence period. As stated above, the best estimate of the recurrence period can thus be found at the crossing of the continuous spectral peaks and the noise regime for small ε (Figure 4a). SRPS may be estimated using the nta_recfreq_en_scan function. The most important parameter is the range of neighbourhood-sizes to be tested (default: 1-100 % of SD of data).

One disadvantage of the method is its lack of temporal resolution. This, however, is important for many scientific fields, where non-stationary data is present. In neuroscience for example, one is often interested in contrasting a baseline at rest with neuronal activity modulated by an internal or external stimulus (Oehrn et al. 2015). A straightforward solution for implementing a time-dependent recurrence amplitude spectrum (TRAS) is to simply divide the time series into sections of equal length and compute the spectrum for each of them.

$$P(T, w_n) = \frac{R(T, w_n)}{\sum_{i=T_{min}}^{T_{max}} R(i, w_n)}, T = T_{min} \dots T_{max}, \qquad ( \ 18 \ )$$

with $w_n$ indicating the nth temporal window of input time series x, with window size s = [$x_n \dots x_n$ +windowsize-1].

A similar approach is typically used for short-time Fourier transform (STFT). Naturally, the choice for the window length determines the longest resolvable recurrence period. In order to smooth the boundaries of each temporal segment, NoLiTiA uses windows with 50 % overlap. In Figure 4c we compare TRAS to short-time Fourier transform (STFT, Figure 4d) using a concatenated signal of three different oscillations each sampled at 1000 Hz: 1) a sinusoid of 33 Hz, 2) a sawtooth signal at 33 Hz and

4) a rectangle signal at 33 Hz. (Figure 4b). Note the occurrence of harmonics in Figure 4d for the non-sinusoidal signals, which are absent in the TRAS. Estimation of TRAS may be invoked by calling the nta_wind_recfreq function. Besides an appropriate neighbourhood-size, it is also important to specify a reasonable time window length (default: 1/10 of data length in samples).

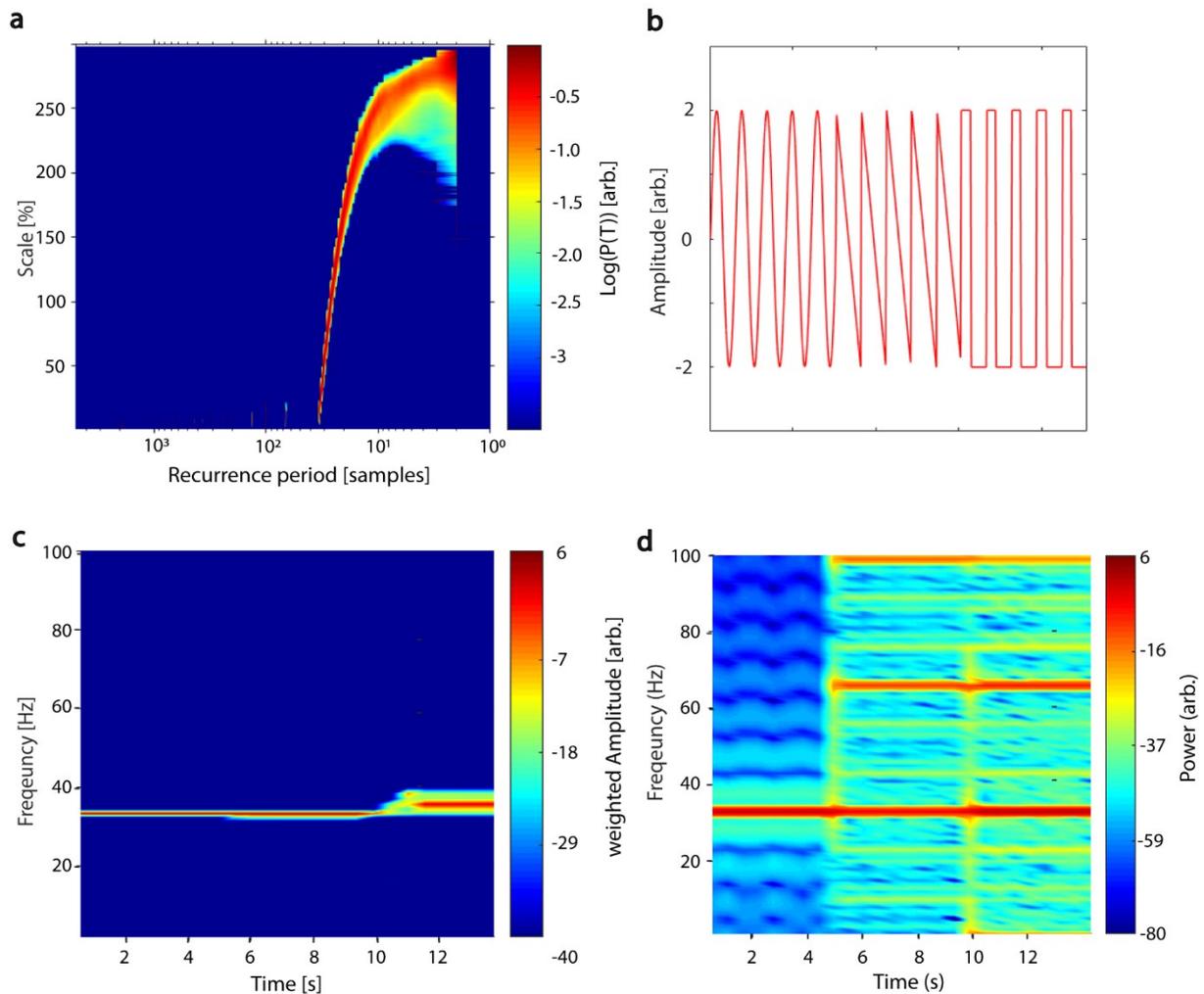

**Figure 4:** SRPS and TRAS of a compound signal. a) SRPS of a 50 s time series with a 3 Hz oscillation, sampled at 100 Hz (+8 % noise relative to STD of raw signal). Note the three regions: 1) for small neighbourhood-sizes SRPS is governed by noise, 2) for large neighbourhoods SRPS is biased towards small periods, 3) at the crossing of the noise level, the true recurrence period of 33 samples can be observed. b) Signal of three concatenated oscillations, each with a duration of five seconds, sampled at 1000 Hz: 1) sinusoid of 33 Hz, 2) sawtooth signal at 33 Hz and 3) rectangle signal at 33 Hz. Note that shorter segments of each oscillatory signal are depicted for illustration purposes. c) TRAS of the compound signal (window length: 1100 ms, overlap 50 %) with weighted amplitudes. d) Short-time Fourier transform of the same signal (window length: 1100 ms, overlap: 50 %). Figure A), C) and D) adapted from Weber and Oehrn (2021).

*Information Theoretic Measures*

One possibility to study complex systems is by analysing its distributed computing capabilities (Fernández and Solé 2006; Lizier 2012). Distributed computing describes a systems information sharing and processing routines. Especially in neuroscience, information theoretic measures have witnessed a surge of interest over the last decade (Dimitrov et al. 2011) and were e.g. used to study neural coding (Quiroga and Panzeri 2009) or communication (Weber et al. 2020a). Measures from information theory quantify the information content of variables or equivalently the amount of uncertainty reduced when measuring the outcome of a random event. The most basic information theoretic measure is the Shannon entropy H (Shannon and Weaver 1949):

$$H(\mathrm{X}) = -\sum_{\mathrm{x}} \mathrm{p_X}(X = x) \log_a \mathrm{p_X}(X = x) \qquad (\ 19\ )$$

or for continuous variables the differential entropy:

$$H(\mathrm{X}) = -\int \mathrm{p(x)} \log_a \mathrm{p(x)dx} \qquad (\ 20\ )$$

where p(x) indicates the probability of variable X taking the value x.

For most information theoretic measures, two different estimators are supplied:

1) an estimator using a simple binning approach (Cover and Joy 1991)

2) a nearest neighbours-based estimator (Kozachenko and Leonenko 1987), with the latter being more computational demanding but unbiased for mutual information.

Most of the functions provided offer an average, as well as a time-resolved local variant of the information theoretic measures (Lizier 2012).

Based on the concept of Shannon entropy, the mutual information determines the shared information content between two processes X and Y (Cover and Joy 1991, Figure 5):

$$I(X;Y) = H(X) - H(X,Y) + H(Y)$$ ( 21 )

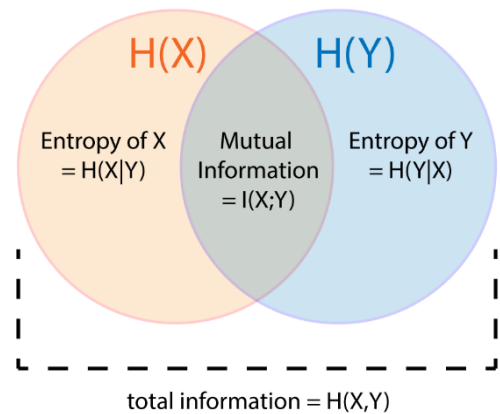

**Figure 5:** Relationship of mutual information to Shannon entropy.

A derived concept is the auto-mutual information between a signal X and a time shifted copy of itself. The auto-mutual information may be used to estimate the optimal delay τ for reconstructing the phase-space (Fraser and Swinney 1986). It is used as a subroutine for the function nta_optimize_embedding when specifiying the input parameter cfg.method as "deterministic" (see section Phase-Space Reconstruction).

By calculating the mutual information between the present and the immediate past state of a process X, one can estimate the active information storage (AIS). AIS quantifies how much information is currently in use for computing the next state (Lizier 2012):

$$AIS(X) = I(\boldsymbol{X}_{t-1}^{d_x}, X_t), \qquad\qquad ( 22 )$$

As the estimation of all entropy-based measures strongly depends on the estimation of the involved probability functions, the number of bins (for binning algorithms, default: 0) and the number of neighbours (for the nearest neighbours-based algorithms, default: 4) should be specified. An optional optimization algorithm for selecting the number of bins for the estimation of probabilities is implemented using the Freedman-Diaconis rule (Freedman, David, and Persi Diaconis 1981).

For a complete list of functions, implemented methods and parameters please refer to the manual in the supplement.

## 3. General Workflow

Depending on the degree of needed flexibility, experience in programming and ease of use, NoLiTiA offers three different ways to analyse data: 1) a graphical user interface (GUI), 2) a batch-editor and 3) custom-made MATLAB-scripts (Figure 6).

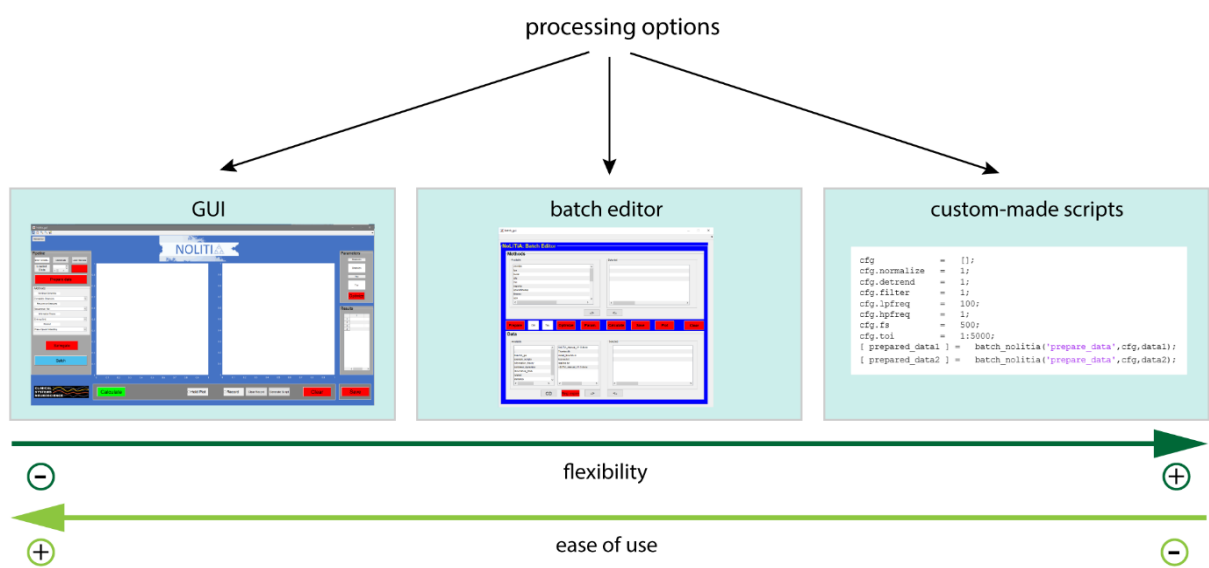

**Figure 6:** Processing options.

Despite slight differences, all three pathways share the same basic processing pipeline. Depending on the user's analysis pathway, data may be loaded either from a file or directly from workspace (GUI). NoLiTiA offers the possibility to pre-process data. The user may specify a time-region of interest, to subtract linear trends, to normalize and/or to filter data (for filtering, the signal processing toolbox™ (The Mathworks, inc.) needs to be installed). The user may choose methods from the three topics: 1) nonlinear dynamics, 2) recurrence-based methods, 3) information theoretic measures (Figure 1). Most of the methods necessitate to define embedding parameters (dim=dimension, τ=embedding delay) for phase-space reconstruction. The three pathways offer either to define them *ad-hoc* by the user or to optimize them based on two different approaches. The optimization procedure should be chosen depending on whether data was generated by a deterministic or a stochastic process (see section Phase-Space Reconstruction).

Every method has at least one parameter, which may be specified by the user. If a parameter is left unspecified, the default value is applied (see the manual in the Supplement for a list of all parameters including default values).

Depending on the analysis pathway, results are either plotted automatically (GUI), or the user may optionally choose to do so (batch and custom-made scripts).

The graphical user interface (GUI) is intended to be the most beginner's friendly option for analysis. The GUI can be invoked by typing nolitia_gui in the command window (Figure 7a). The interface is composed of four main regions: on the left side, the user loads the input data, chooses whether and how to pre-process, specifies analysis methods, generates surrogate data and enters batch-mode. On the right side, the user may enter embedding parameters (dim and τ) *ad-hoc* or choose to optimize them using two different approaches. Method-specific results are displayed in a designated table. In the middle of the interface, two axes display method-specific figures after calculation. By clicking the 'Hold Plot' radar button the user may superimpose results of subsequent analyses. By toggling the 'Record'-button, all main steps and commands done by the user are saved in a queue. Clicking the 'Generate Script'-button automatically generates a MATLAB-script of all operations saved in the queue.

The batch-editor is intended to be a compromise between the accessibility of the GUI and the flexibility of custom-made scripts. In contrast to the GUI, it allows for a semi-automatized stacked analysis of multiple datasets and methods. The batch-editor can be invoked by either clicking on the 'batch' button in the GUI (Figure 7a) or by typing batch_gui in the command line of MATLAB. The batch-editor is composed of three main parts. In the bottom half, the user loads datasets and chooses which data to analyse. In the top half, the user chooses which methods to use for analysis. In the center, linearly arranged buttons guide the user through the analysis pipeline (Figure 7b).

Either clicking on the 'Plot'-button or typing plot_batch_gui in the command line loads the Plotting-Tool (Figure 7c). The plotting-tool is intended to be used for displaying results generated by the batch-editor. If the tool is loaded by clicking the Plot-button after computation in the batch-editor, the results of the first data set are automatically loaded into the plotting-tool. Alternatively, the user may load saved data by clicking the 'Load'-button. Available methods are represented by a hierarchical tree structure, which can be unfolded by clicking on Methods in the method panel. Methods which can be plotted using the Plot-button are indicated by a green arrow next to its name. Results are plotted in the centre axis.

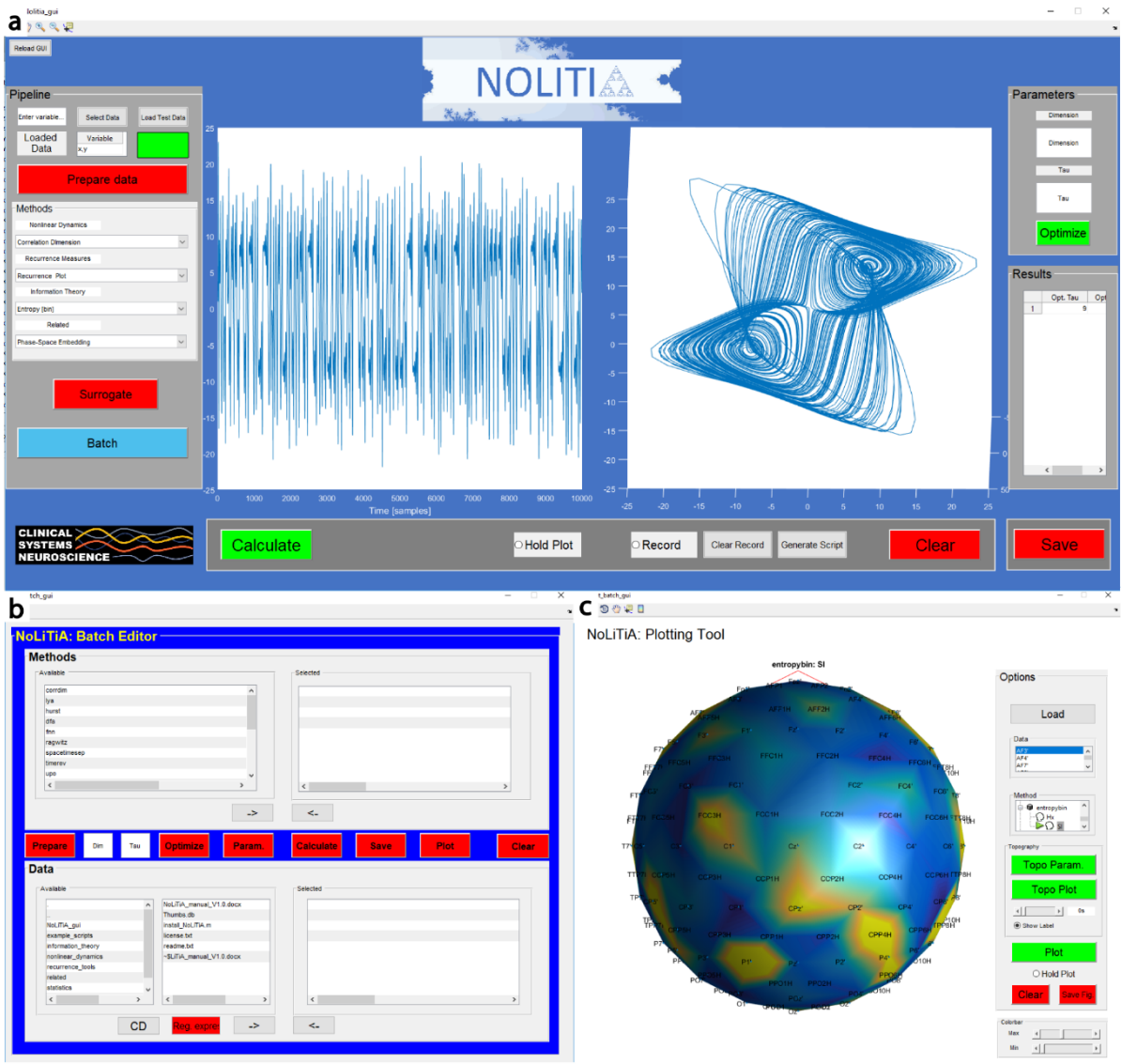

**Figure 7:** Interfaces of NoLiTiA. a) GUI depicting a phase-space reconstruction of the Lorenz attractor, b) batch-editor, c) plotting-tool showing a topographic representation of Shannon entropies estimated from artificial EEG data.

The plotting tool may be used to analyse electroencephalographic (EEG)- data, as each time series may represent one recorded channel. The plotting-tool allows for a topographical representation of results per channel. By clicking on the red 'Topo Param.'-button the user is prompted to select an electrode-position file containing channel names and electrode positions. Additionally, the user may specify specific electrodes to plot, as well as a sampling frequency. Finally, after selecting a method from the list, the topographic plot is displayed after clicking the 'Topo Plot'-button (Figure 7c). Methods, which can be plotted using the 'Topo Plot' -button are indicated by a head shape next to its name. Methods with a green arrow, as well as a head shape symbol next to it may be plotted time resolved. This can be done by first clicking the 'Topo Plot'-button and then using the slider below to scroll forward or backward in time. Supported function are listed in the Supplement.

The final option to analyse data is by using custom-made scripts. This option offers the highest flexibility but requires some experience in MATLAB-scripting. All core functions are designed in the same way. Following the naming convention of other MATLAB toolboxes like FieldTrip (Oostenveld et al. 2011) or SPM (The Wellcome Dept. of Imaging Neuroscience, London; www.fil.ion.ucl.ac.uk/spm) most of NoLiTiAs functions start with a prefix (nta_). The user must provide up to two datasets (depending on whether the method is uni- or bivariate) and a configuration structure cfg containing method-specific parameters. An example script is shown below. Default values exist for all parameters. For a complete list of parameters per method see the manual in the Supplement. For beginners the GUI offers the possibility to record processing steps made in the GUI and to save them as a MATLAB script. Differences between datasets, e.g. measured conditions in cognitive neuroscience may be statistically tested using non-parametric procedures, i.e. a Monte-Carlo simulation (nta_nolitia_monte) or cluster-based permutation algorithm (nta_nolitia_cbpa, Oostenveld et al. 2011).

```
Example analysis pipeline

%% Generate data set of test and surrogate data (2*10 subjects/10 trials/3
variables/5981 samples)

generate_test_data

%%%%%%%%%%%%%%%%%%%%%%%%%%%%%%%%%%%%%%%%%%%%%%%%%%%%%%%%%%%%%%%%%%%%%%%%%
%% Analyze data

cfg_a1.Data          =      data;

cfg_a1.methodnames   =      {'nta_entropybin','nta_hurst'};
cfg_a1.verbose       =      0;
[cfg_a1]             =      prepare_wrapper(cfg_a1);
[cfg_a1]             =      generate_cfg(cfg_a1);
[cfg_a1]             =      batch_wrapper(cfg_a1);

cfg_a2.Data          =      data_surr;
cfg_a2.methodnames   =      {'nta_entropybin','nta_hurst'};
cfg_a2.verbose       =      0;
[cfg_a2]             =      prepare_wrapper(cfg_a2);
[cfg_a2]             =      generate_cfg(cfg_a2);
[cfg_a2]             =      batch_wrapper(cfg_a2);

%%%%%%%%%%%%%%%%%%%%%%%%%%%%%%%%%%%%%%%%%%%%%%%%%%%%%%%%%%%%%%%%%%%%%%%%%
%% Monte-Carlo Test
Cfg                  =      [];
cfg.method           =      'nta_entropybin'; %'hurst'
cfg.outputvar        =      'Hx';%'expo';
cfg.numperm          =      1000;
cfg.vars             =      1:3;
cfg.plt              =      1;

[data1,data2]        =      fetch_data(cfg,cfg_a1,cfg_a2);

results_monte        =      nta_nolitia_monte(data1,data2,cfg);

%%%%%%%%%%%%%%%%%%%%%%%%%%%%%%%%%%%%%%%%%%%%%%%%%%%%%%%%%%%%%%%%%%%%%%%%%
%% Cluster-based permutation analysis

cfg=[];
cfg.method           =      'nta_entropybin'; %'hurst'
cfg.outputvar        =      'SI';%'expo';
cfg.numperm          =      1000;
cfg.vars             =      1:3;
cfg.maxcluster       =      'sumt';
cfg.clustalpha       =      0.05;
cfg.plt              =      1;
results_cbpa         =      nta_nolitia_cbpa(data1,data2,cfg);
```

## 4. Validation and Example Applications

*Lorenz Data*

The scripts of the following examples can be found in the Supplement. The first example data set comprises 250 s of the X-component of the Lorenz system (Lorenz 1963):

$$\dot{X} = a(Y - X)$$

$$\dot{Y} = X(b - Z) - Y \qquad\qquad ( \ 23 \ )$$

$$\dot{Z} = XY - cZ$$

with random initial condition and sampled at 40 Hz. The Lorenz system is a classic mathematical model, initially designed to study atmospheric convection. However, it is better known for its well characterized chaotic behavior invoked by using specific parameters. Typical parameters a=10, b=28 and c=8/3 were used in order to invoke chaotic behaviour (Lorenz 1963). To reconstruct the phase-space, the embedding delay τ=8 was chosen according to the first minimum of the auto-mutual information (Figure 8a) and the embedding dimension parameter dim=3 using the false nearest neighbours algorithm (Kennel et al. 1992) over a range of embedding dimensions d=1-9 (Figure 8b). The original as well as the reconstructed attractor are topologically similar (Figure 8c-d). The typical attractor with its two wings can be recognized both in the original Figure 8c) and reconstructed phase-space (Figure 8d). Next, nonlinearity was tested using the time inversion statistic and 1,000 amplitude-adjusted phase randomized surrogates as suggested by Schreiber and Schmitz (Schreiber and Schmitz 1997). As expected for a nonlinear system, the null hypothesis of linearity was rejected at an alpha-level of 5 % (Z-score=-2.19, Figure 9a). The correlation dimension was estimated using the Grassberger-Procaccia algorithm (Grassberger and Procaccia 1983) and the largest Lyapunov exponent using the Kantz algorithm (Kantz 1994). The Theiler-window of temporal correlations was chosen according to the plateau region of the space-time separation plot at about dt=20 (Figure 9b).

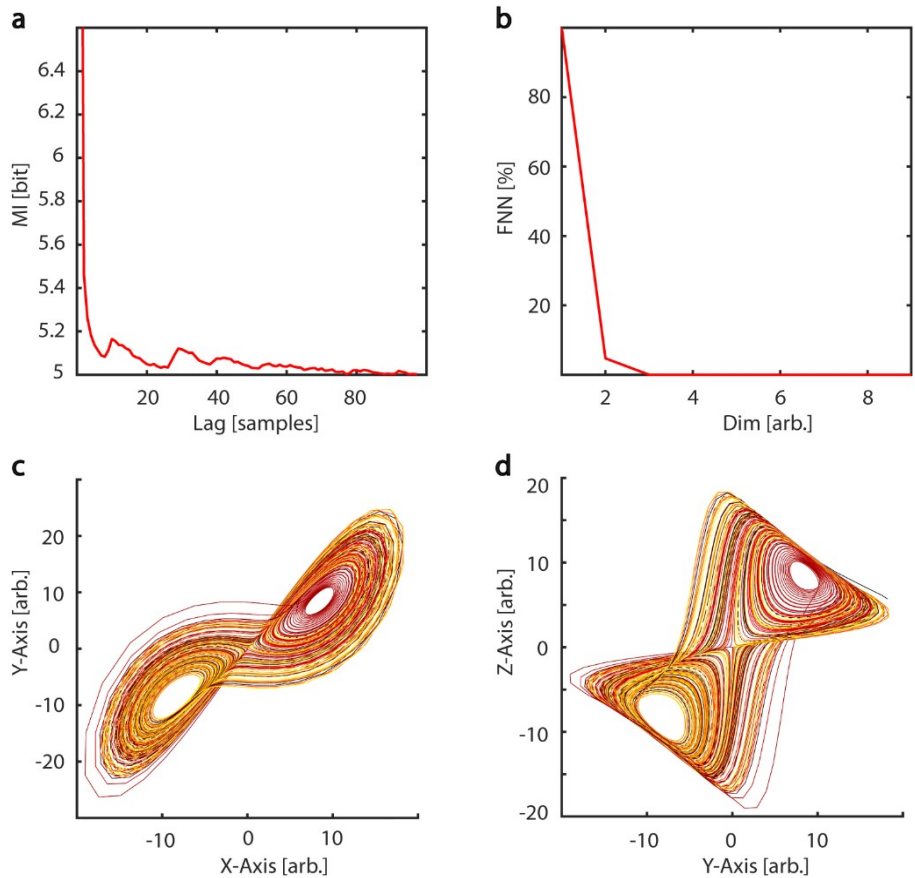

**Figure 8:** Embedding of Lorenz data. a) The embedding delay τ=8 was chosen according to the first minimum of the auto-mutual information function. b) Cao's method of using the false nearest neighbours algorithm revealed d=3 as the optimal embedding dimension. c) - d) Original and reconstructed Lorenz attractor. MI: mutual information, FNN: false nearest neighbours, dim: dimension.

Results for correlation dimension and Lyapunov exponent both asymptotically reach their analytically calculated values (Figure 9c-d, D2=2.05, λ=0.906, Grassberger and Procaccia 1983; Sprott 2003). Finally, a spatially (SRPS), as well as a temporally (TRAS) resolved recurrence periodogram was created. It quantifies the probability of recurrences in phase-space as a function of recurrence periods and either spatial scales (equation 18), i.e. neighbourhood-sizes or time using overlapping temporal windows (equation 19). For the example analysis, spatial scales ranging from 1-100 % of the standard deviation and temporal windows of 400 samples with 50 % overlap were used. Results are summarized in Figure 10a-b and reveal intermittent recurrences with periods at multiples of 24 samples.

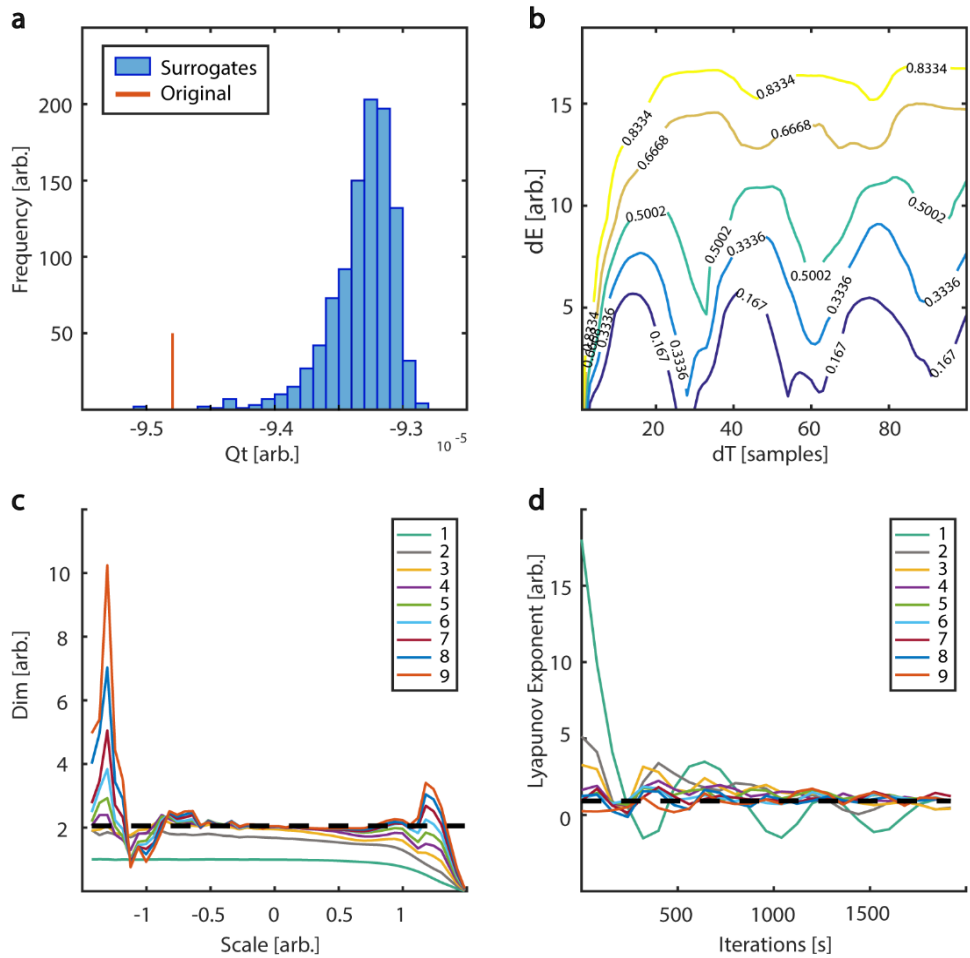

**Figure 9:** Tests for nonlinearity and determinism. a) Frequency distribution of time reversibility statistic of 1,000 surrogates in comparison to original data. The null hypothesis of linearity could be rejected at an alpha level of 5 % (Z-score=-2.19). b) Space-time separation plot revealed a plateau at ~dT=20 which was subsequently used as Theiler-window. Coloured lines indicate the proportion of states indicated by the superimposed numbers. c) Estimation of correlation dimension for embedding dimension d=1-9. Analysis revealed a plateau region slightly above two for embedding dimension 3-9 between spatial scales of ~-0.7 to 0.7, which is in line with the analytic value (slashed line). d) Estimation of largest Lyapunov exponent for embedding dimensions d=1-9. In accordance with the analytic value (slashed line=0.906) estimation revealed a maximum Lyapunov exponent of ~0.9. dE: spatial separation, dT: temporal separation, dim: dimension, Scale: logarithmic neighbourhood-size in phase-space.

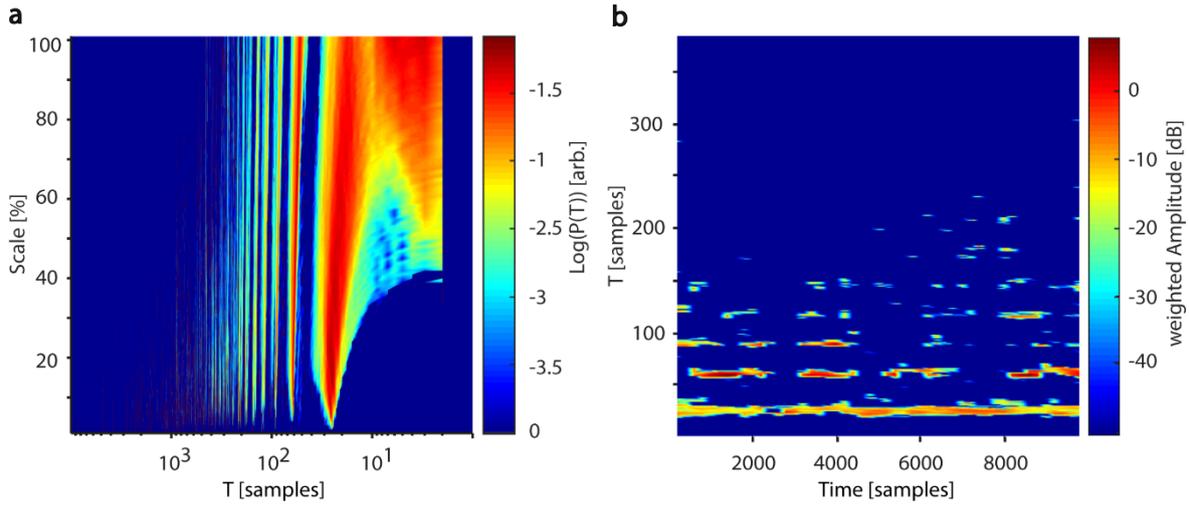

**Figure 10:** Recurrence periodograms of Lorenz data. a) Recurrence probabilities as a function of recurrence periods and spatial scale. A high probability of recurrence can be seen at multiples of T~24 samples (~1.67 Hz). b) Time-resolved recurrence periodogram with window size w=400 samples at 50 % overlap. Similar to a) multiples of a fundamental period of ~24 samples can be detected. The windowing approach reveals intermittent notches in the spectrum (e.g. at ~2000 samples), which are concurrent with the Lorenz attractor reinjecting its trajectory to the center of its "wings" (see Figure 8c-d). Scale: Neighbourhood-size in % of standard deviations of Lorenz data, T: recurrence period, P(T): probability of recurrence at period T.

*Logistic Map*

In order to validate the estimation of period one unstable periodic orbits, 100 iterates of the logistic map (Ausloss and Dirickx 2006) were created with random initial condition and a parameter a=3.92 to invoke chaotic dynamics:

$$\text{f}: \quad x_{n+1} = ax_n(1 - x_n) \quad\quad\quad ( \ 24 \ )$$

The transformation procedure was performed 500 times. The histogram of the original two-dimensional phase-space is depicted in Figure 11a, while Figure 11b shows the phase-space after the transformation procedure. In Figure 11c, the probability of a phase-space occurrence is displayed as a function of X-axis coordinates of the first diagonal, both, for the original and transformed phase-space. After the transformation procedure, points of the phase space have been accumulated at approximately 0.75. A permutation test with 1,000 amplitude-adjusted phase randomized surrogates was applied to test for significance of the correctly detected fixpoint at 0.75 at an alpha-level of 5 % (Z-score=15.32, Figure 11d). The results are in accordance with previously reported results (So et al. 1996).

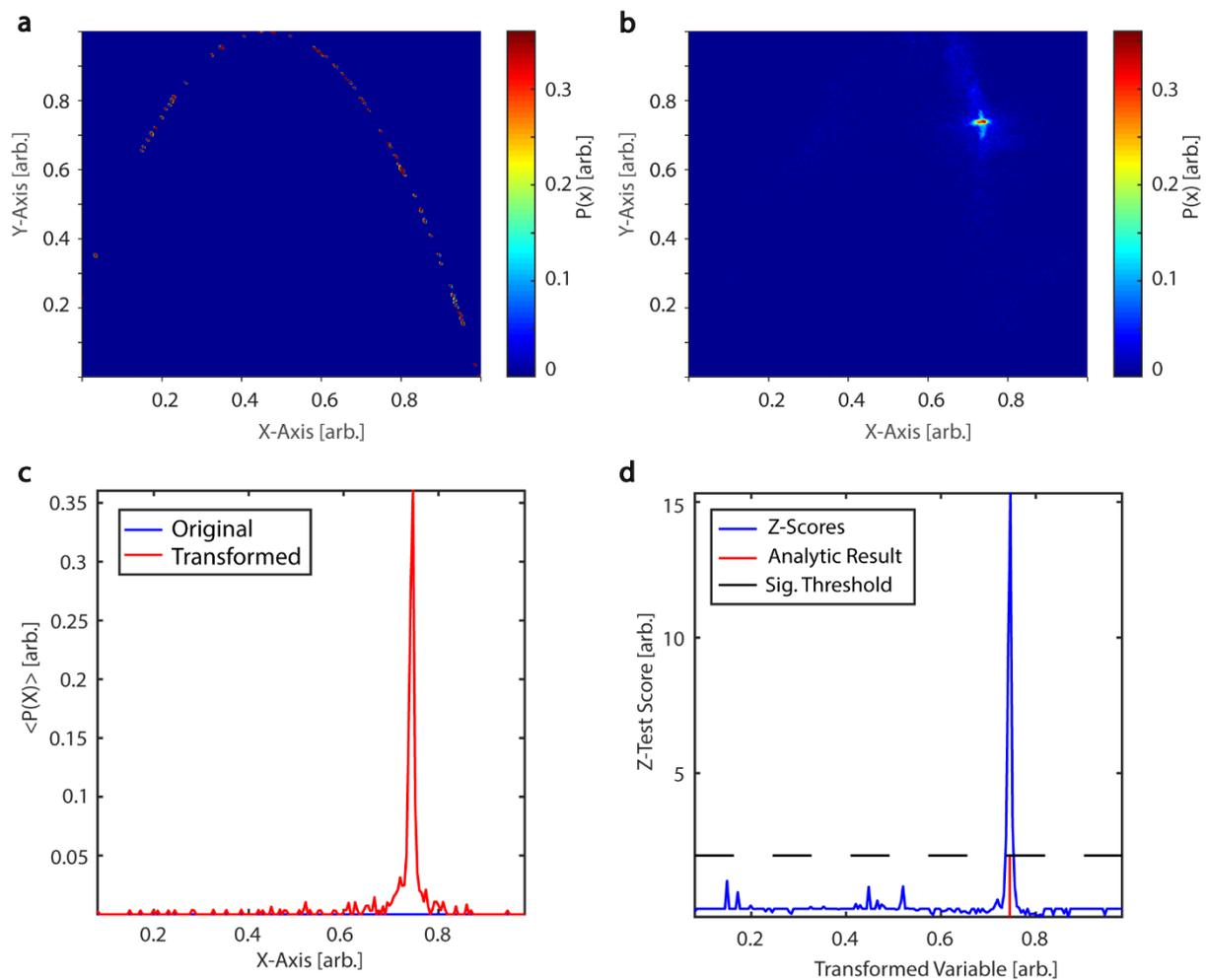

**Figure 11:** Periodic orbit transform of the logistic map. a) Probability map of the original phase-space. b) Probability map after periodic orbit transform. c) Diagonals of the original and transformed probability map.

After transformation, states in its vicinity are mapped onto the fixpoint at 0.745. d) Surrogate test reveals a significant fixpoint at 0.746. P(x): probability of state x, <P(x)>: mean probability over 500 transformations.

*Gaussian Distributions*

Information theoretic measures were validated using Gaussian distributions of unit variance and a covariance of 0.9. Figure 12a-b summarize estimation results in comparison to analytic values for the binning (Cover and Joy 1991) and neighbourhood-based estimator (Kozachenko and Leonenko 1987) of Shannon entropy and differential entropy, respectively. Entropy can be calculated analytically for Gaussian distributions according to:

$$H = \frac{1}{2}\log_2(2\pi e\sigma^2) \qquad\qquad (\ 25\ )$$

Similarly, Figure 12c-d display results for mutual information estimation using a binning (Cover and Joy 1991) and a nearest neighbour-based approach (Kraskov et al. 2004). Mutual information for a bivariate Gaussian distribution is given by:

$$I(x;y) = -\frac{1}{2}\log_2(1 - \frac{\sigma_{xy}^2}{\sigma_y\sigma_y}) \qquad\qquad (\ 26\ )$$

The results of the implemented algorithms are in accordance with previously published results (Kraskov et al. 2004). Both, the binning and the nearest neighbours estimator of entropy asymptotically reach a stable difference dH with respect to the analytical values at approximately $10^{-2}$. Locally, for few neighbours, the nearest neighbours estimator even reaches a difference of approximately $10^{-4}$. Comparison of mutual information estimators with analytical values shows, that the nearest neighbours estimator is reliable for a broader parameter range, than the binning estimator. While the binning estimator overestimates mutual information for increasing number of bins, the nearest neighbours estimator underestimates it for a number of neighbours larger than 100.

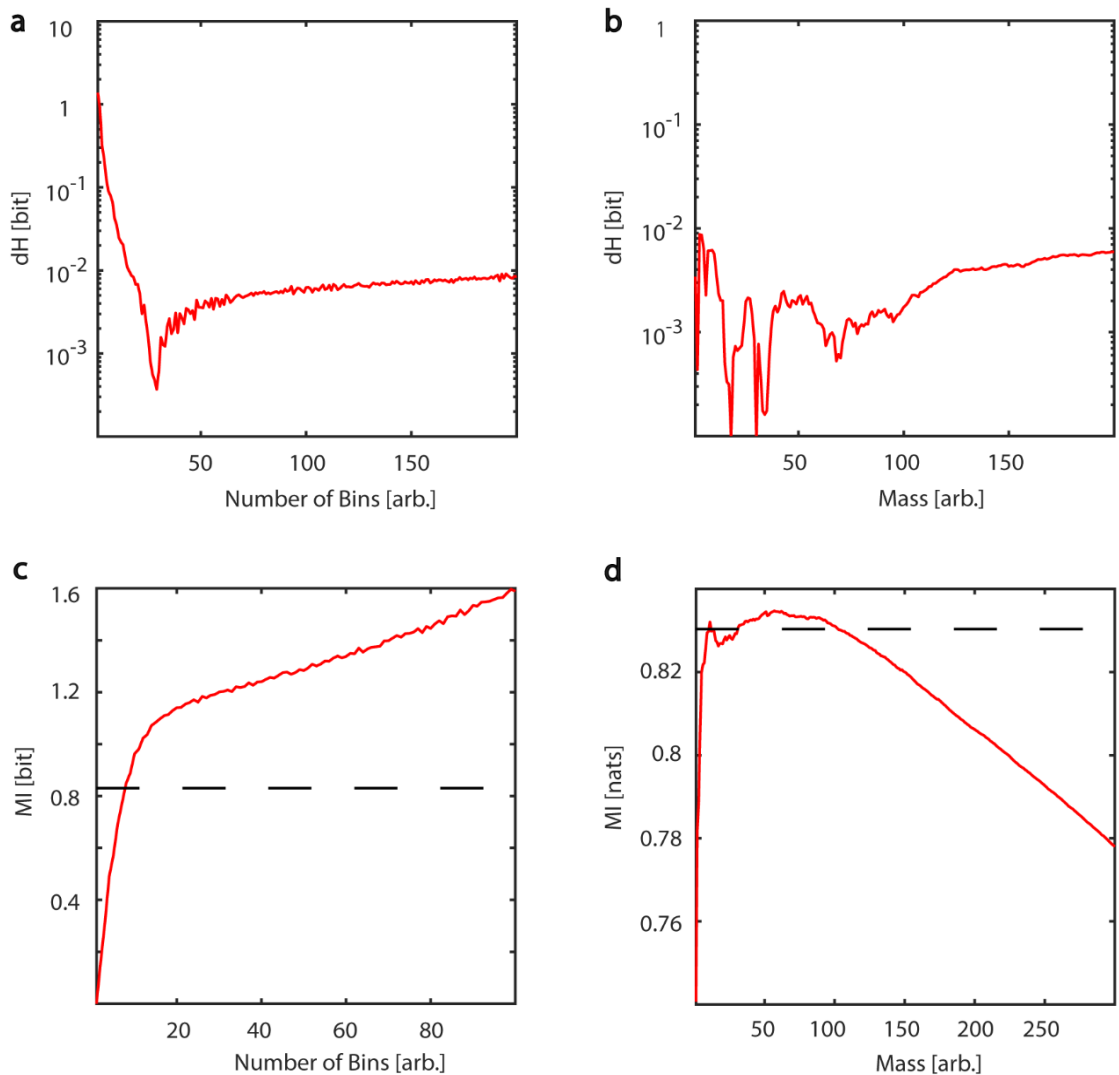

**Figure 12:** Basic information theoretic measures. a) Difference between analytic and estimated Shannon entropy as a function of number of bins. b) Difference between analytic and estimated differential entropy as a function of mass using the Kozachenko-Leonenko estimator. c) Mutual information as a function of number of bins. d) Mutual information as a function of mass, using Kraskov's estimator. Slashed line: analytic value, dH: entropy difference between analytic and estimated value, MI: mutual information, mass: number of neighbours used for estimation of probability densities.

*Patient Data*

In a fourth example ~4 s (10,000 samples) of intraoperatively recorded electromyographic activity (EMG) of a Parkinson's disease patient at rest during tremor activity was analysed. The aim was to characterize tremor activity with respect to frequency content and whether it can be described to be generated by either a linear stochastic, nonlinear stochastic or nonlinear deterministic process. The data was recorded from the right extensor digitorum communis (EDC) using the INOMED ISIS MER-system (INOMED Corp., Teningen, Germany) and sampled at a frequency of 2456 Hz. The patient signed written informed consent prior to publication. Data collection was approved by the local ethics committee and was in accordance with the Declaration of Helsinki. The same data was previously published by Florin et al. (2010).

Using NoLiTiA, data was first trend corrected and normalized to zero mean and unit variance. Next, a fourth order, bidirectional, 20 Hz high-pass Butterworth filter was applied to remove activity unrelated to EMG. Since no a priori assumption could be made regarding the question whether the generating process was deterministic or stochastic, the optimal embedding dimension was estimated using the false nearest neighbours algorithm and the Ragwitz criterion and results compared (Figure 13b-c). The embedding delay was co-optimized by the Ragwitz criterion and additionally determined using the first minimum of the auto-mutual information function (Figure 13a, c).

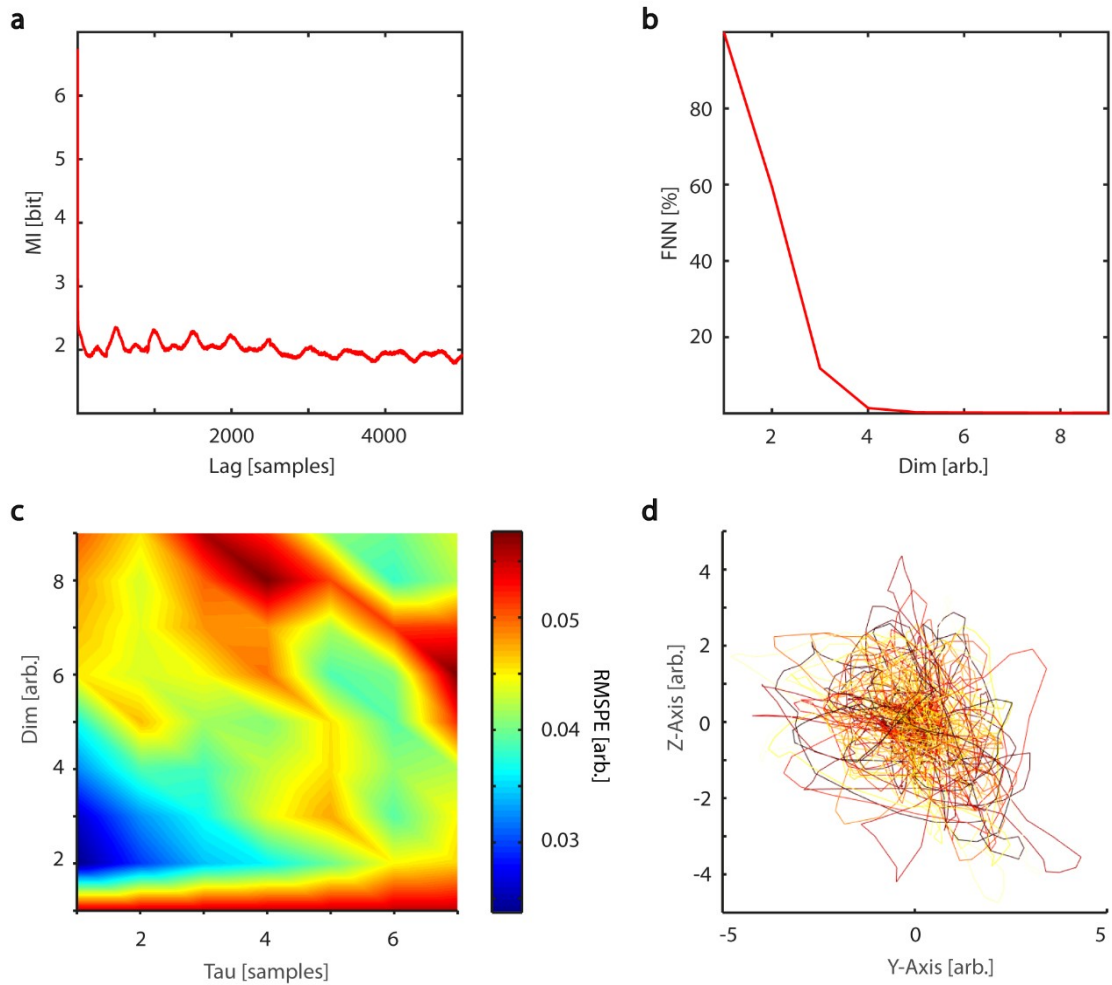

**Figure 13:** Embedding procedure of EMG data. a) The auto-mutual information revealed a first-minimum at 14 samples, which was subsequently used as embedding delay. b) The method of false nearest neighbours led to an optimal embedding dimension of 5. c) Additionally, the Ragwitz criterion was used to co-optimize embedding dimension and delay. As expected[17], the procedure suggested a much smaller delay and embedding dimension with 1 and 3, respectively. d) Embedding of EMG data with $\tau=14$ and $d=3$ for illustration purposes. MI: mutual information, FNN: false nearest neighbours, dim: dimension, RMSPE: root-mean-square prediction error.

Nonlinearity was tested using a surrogate test and the time-inversion statistic as suggested by Schreiber and Schmitz (Schreiber and Schmitz 1997). 1,000 amplitude-adjusted phase randomized surrogates were created. The null hypothesis of linearity was rejected at an alpha-level of 5 % Figure 14a.

Next, determinism was tested by estimating the correlation dimension and maximum Lyapunov exponent over a range of embedding dimensions d=1-9. If a stable embedding dimension invariant estimate of either statistic could be detected the null hypothesis of stochasticity could be rejected. For the Theiler-window of temporal correlations, twice the autocorrelation time was excluded. Additionally, a space-time separation plot was produced to verify the choice. Results are summarized in Figure 14c-d. While the null hypothesis of linearity could be rejected, neither an embedding dimension invariant scaling range for the correlation dimension nor exponential divergence of neighbouring states, i.e. a positive maximum Lyapunov exponent could be detected.

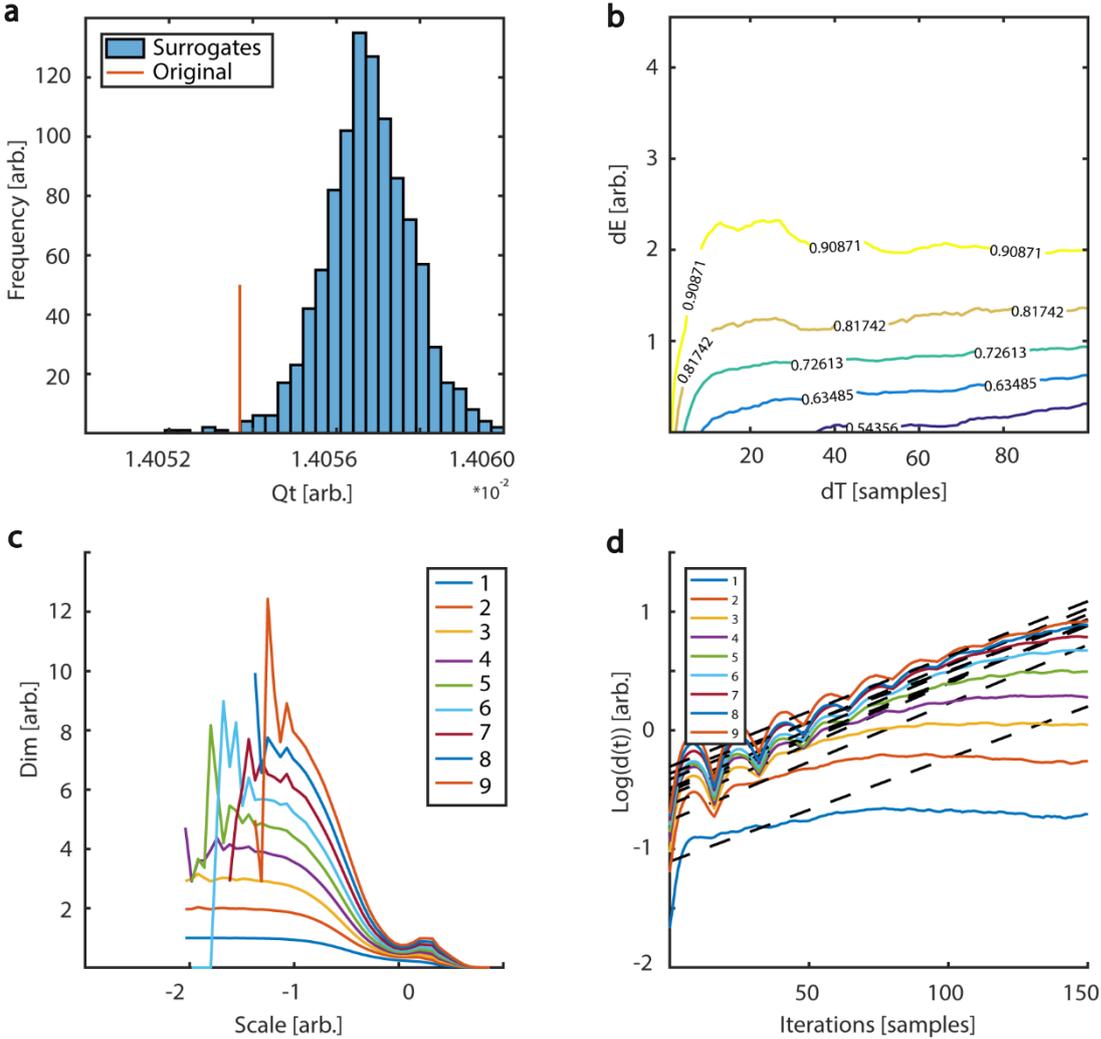

**Figure 14:** Testing for nonlinearity and determinism. Distribution of 1,000 surrogates for nonlinearity test. Following the surrogate test, which employed a time inversion statistic, the null hypothesis of linearity could be rejected at an alpha-level of 5 % (Z-score=2.85). b) The space-time separation plot shows a plateau region at ~20 samples which was subsequently used as Theiler-window. Colored lines indicate proportion of data points as indexed by superimposed numbers. c) Estimation of correlation dimension for embedding dimensions d=1-9 as a function of logarithmic spatial scales. Neither a clear scaling region, nor any convergence for subsequent embedding dimensions could be detected. d) Average distances of neighbouring states in phase-space as a function of temporal iterations are plotted for embedding dimensions d=1-9. Neither a clear exponential expansion rate nor any convergence for subsequent embedding dimensions could be detected. Qt: time-inversion statistic, dE: spatial distance, dT: temporal distance, dim: dimension, d(t): average distance of next neighbours after t iterations.

Finally, the frequency content of the EMG was analysed by calculating 1) recurrence periods time independent over a range of spatial scales (SRPS) $\varepsilon$=1-100 % of the standard deviation of EMG data and 2) by estimating recurrence periods time-resolved using overlapping windows of 0.5 s and a fixed spatial scale of $\varepsilon$=10 % (TRAS). Temporally and spatially resolved recurrence period analysis revealed prominent recurrence periods at multiples of ~500 samples, i.e. 203 ms (4.9 Hz), probably indicating tremor activity (Figure 15a-b). In conclusion, the EMG-data of the Parkinson's patients can be characterized as generated by a nonlinear stochastic oscillator at approximately 5 Hz.

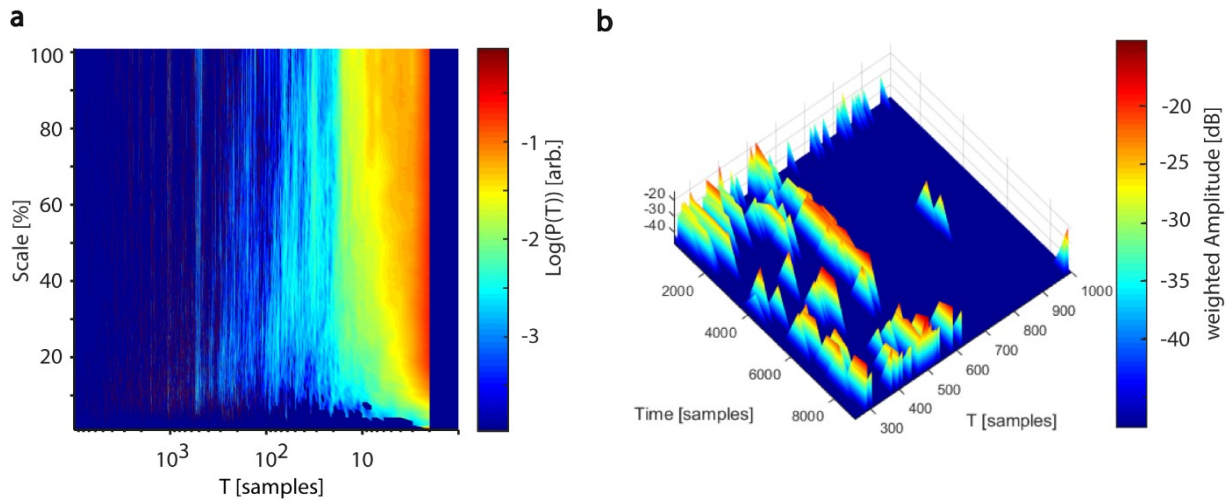

**Figure 15:** Analysis of recurrence periods. a) Logarithmic recurrence period probabilities as a function of recurrence periods T and spatial scale (SRPS). b) Logarithmic recurrence period probabilities as a function of recurrence periods T and time using windows of 1200 samples with 50 % overlap (TRAS). In both plots prominent recurrence periods are visible at multiples of ~500 samples (~4.9 Hz).

## 5. Summary and Conclusion

In this study a free, open-source toolbox for nonlinear time series analysis, including several established and novel measures, was presented. The implementation of methods, ranging from dynamical systems theory, recurrence analysis to information theory were validated on artificial data. An example application was given for electrophysiological data of a Parkinson's disease patient. In comparison to similar toolboxes (Hegger et al. 1999; Donges et al. 2015; Lizier 2014), NoLiTiA offers three advantages: 1) By combining methods and algorithms from three broad fields of complexity theory, the package covers a wide range of applications within the same framework, e.g. ranging from characterizing stochastic or deterministic systems to analysing oscillatory phenomena. 2) The GUI and batch-editor offer accessibility of nonlinear methods even for scientists unexperienced in programming. 3) Aside from established methods, two recently proposed measures for quantification of oscillatory activity were implemented. SRPS and TRAS each aim to extend the recurrence period density estimation (Little et al. 2007) by estimation of the recurrence amplitude, as well as a spatial and temporal dimension, respectively. SRPS may be applied to choose an optimal neighbourhood-size for a subsequent time-resolved analysis. Similar to STFT, TRAS allows to analyse oscillatory activity as a function of time. However, an advantage of TRAS is its applicability on nonlinear signals, i.e. non-sinusoidal oscillations (Weber and Oehrn 2021). Analysis of such signals by means of STFT and related measures lead to the appearance of spurious harmonics in the resulting spectra. As the shape of the oscillations is irrelevant for TRAS, as long as similar states recur, even highly asymmetric signals, like sawtooth oscillations, generate sharp peaks in the TRAS. Preliminary code of NoLiTiA has already been successfully applied in a diverse range of previous studies ranging from a normative study for action pictures and naming latencies to analysis of EEG and invasive recordings in Parkinson's patients (Busch et al. 2021; Loehrer et al. 2021; Oehrn et al. 2021; Weber et al. 2020b; Weber and Oehrn 2021). Future updates will continuously add recent methods to offer the most up-to-date software library for nonlinear data analysis.

## 6. Implementation

With the exception of the low-level function to estimate distances in phase-space (nta_neighsearch), all functions are implemented and validated in MATLAB 2016b. To accelerate computation, the former is implemented in C as a Mex-function and provided as a compiled Mex64 file for 64bit operating systems. To guarantee ease of use, most functions are implemented in the same way, demanding up to two input data sets and one configuration structure. The output is a single "results" structure consisting of fields for provided input parameters, as well as estimation results. Methods were validated using analytic solutions of the Lorenz-system, logistic map and standard distributions, i.e. uniform and Gaussian distributions. The toolbox is open-source and distributed under 2-clause BSD license.

## 7. Availability

The toolbox is readily available to download at http://nolitia.com. The supplementary files may be downloaded at http://nolitia.com/Supplement.zip.

## 8. Acknowledgments

The authors would like to thank Dr. Michael von Papen for discussions during the implementation of methods, Prof. Dr. Esther Florin for providing an example data set, Prof. Dr. Lars Timmermann for guidance during the planning of the study and Jonas Gerards for testing the toolbox and its functions.